# Prioritizing Risk Factors in Media Entrepreneurship on Social Networks: Hybrid Fuzzy Z-Number Approaches for Strategic Budget Allocation and Risk Management in Advertising Construction Campaigns

Ahmad Gholizadeh Lonbar[1,*], Hamidreza Hassanzadeh[2], Fahimeh Asgari[3], Elham Khamoushi[4]
Hajar Kazemi Naeini[5] , Roya Shomali[6], Saeed Asadi[7]

*[1] Department of Civil, Construction, and Environmental Engineering, University of Alabama, Tuscaloosa, AL, USA*
*[2] Department of Environment and Energy, Science and Research Branch, Islamic Azad University, Tehran*
*[3]Department of G. Brint Ryan College of Business, University of North Texas, Texas, USA*
*[4]Department of English, Iowa state university, Iowa, USA*
*[5]Department of Civil Engineering, The University of Texas at Arlington, Arlington, Texas, USA*
*[6]Department of Information Systems, Statistics and Management Science, The University of Alabama, Tuscaloosa, USA*
*[7]Department of Civil Engineering, The University of Texas at Arlington, Arlington, Texas, USA*
**Corresponding author: Agholizadehlonbar@crimson.ua.edu*

**Abstract**

The proliferation of complex online media has accelerated the process of ideology. The formation process can also be influenced by interested people through advertising channels, the media through which messages are distributed to the target audience. There is also a trade-off between using cheap channels versus using expensive but effective channels. These rival forces make it difficult to prioritize the optimal allocation of funds. There are also important technical challenges to solving this problem, because describing the optimal budget allocation between channels over a period of time requires describing the optimal finite vector structure of the controls on the chart. To achieve the most productivity in marketing, you need to determine how a given budget can be allocated to all channels to the actions that are done for business purposes such as revenue, ROI (return on investment), rate Maximize growth and the like. Therefore, the strategy of the media budget allocation model is primarily as an exercise focused on cost and achieving the goal, by identifying a specific framework for a media program. Numerous researchers in the field of optimizing the achievement and frequency of media selection models to help superior planning decisions in the face of the complexity and large amount of information available, so in this study to provide a planning model We do the math to program the media for advertising construction campaigns, using the media mix model. Also, In this case, a decision-making strategy centered on the FMEA is used to identify and prioritize risk factors financial of the media system in companies. Nevertheless, because this strategy has some significant limitations, this research has presented a decision-making approach depending on Z-number theory. For tackle some of the RPN score's drawbacks, the suggested decision-making methodology combines the Z-SWARA and Z-WASPAS techniques with the FMEA method.

# 1. Introduction

With the rapid development of e-commerce, social media has created new opportunities for both consumers and companies, so that marketing channels have increased today. In addition to the traditional advertising through television, radio, direct mail packages, magazines, newspapers, outdoor billboards and the like, digital advertising, online shows, video, communities and email in recent years to target Accurate and fast tracking of functions has attracted a lot of attention. This has become one of the main factors of the consumer revolution. Companies can analyze data on the web and social media to gain valuable insights into consumer demands [1]. However, the extent of their impact is a separate issue. The importance of this decision has grown with the increase in resources allocated to these efforts: in 2017, more than $ 1 trillion was spent on marketing worldwide [2, 3], while $ 8.9 billion was spent on advertising in the 2016 US elections [4]. Social media and web analytics can help companies gauge the impact of their ads and how they send messages to consumers. Companies can also turn to social media analytics to learn more about their consumers. Today, every business, in order to maintain and improve its position in competitive markets, regardless of type, size or industry, must produce media content for its target audience [6, 5]. Therefore, the marketing managers of organizations are always trying to maintain and expand their customers by planning the media and spending budgets in this field. Importantly, the amount of media content that most businesses produce and share on a regular basis to stay in touch with their audience can be confusing in terms of tracking, planning, organizing, distributing and analyzing. The best way to deal with these issues is to do media planning. Media planning is a process that determines where, how and for how long to use advertising media to achieve the goals set in the media strategy [7]. These media include digital media (websites, social networks, and apps), environmental media (billboards, billboards, cinemas, vehicles, and shopping malls), radio and television.

Media planning is the process by which marketers determine where, when, and how often to advertise to maximize engagement and return on investment. This media plan may split the cost of advertising and resources between different online and offline channels such as broadcast, print, paid advertising, video advertising or native content [3]. In today's competitive marketing environment, marketers need to serve consumers with the right message, at the right time and in the right channel to see contributions. An effective media plan leads to a set of advertising opportunities that target specific audiences and fit the organization's marketing budget. Researchers in recent years have found it necessary to study the media in a critical and coherent way, because the media has occupied a central position in our cultural and political life. In fact, we know about the world beyond the direct experiences we receive through the media. In fact, sometimes the media dominates many aspects of our society and individual consciousness, which causes the media to influence the unconscious aspects of our lives with their power. Therefore, using media planning, the audience can be best targeted and communicated with [10]. In fact, due to the spread of technology in the world, media planning should be done according to the taste of the audience and in line with the progress of technology. Therefore, in this research, we present a mathematical planning model for media planning for advertising construction campaigns.

## 2-Media planning

One simple dimension of media planning is choosing the right media models based on the message and the target audience of an advertising program, but the more complex dimension is when, where and how with how much repetition the message should reach what percentage of the target audience so that we Get closer to your brand marketing communication goals. The results of the Internet media planning model offer a new approach that can increase the efficiency of Internet advertising. An advertiser should consider as many sites as possible, because advertising widely on a variety of sites minimizes wasteful costs [11]. If a campaign focuses on only two or three sites, the advertiser must extend the duration of those selected sites. And it significantly reduces the effectiveness of the campaign, because it is more expensive to get the audience to understand the advertising message. Since the pricing method for online advertising is based on the number of page views, buyers should first consider the issue of efficiency [12].

### *2-1-Media budget allocation model*

Therefore, the strategy of the media budget allocation model is primarily as an exercise focused on cost and achieving the goal, by identifying a specific framework for a media program. Numerous researchers in the field of optimizing the availability and frequency of media selection models to help superior planning decisions in the face of the complexity and vast amount of information available, which we also in this study to address the importance of media budgeting. In this research, we will provide a model based on statistical methods and machine learning. The items mentioned in this section show the importance and necessity of this research

# 2. Related works

Patterson and Yogef (2020) proposed a system of budget allocation and optimization in the management and payment system of multichannel multimedia scans. They provided a way to manage and integrate resources for a marketing / advertising scandal in which it allocated resource budgets and provided customer service and processing performance. The current technology provides a platform for the customer retention and customer management system to automatically optimize and allocate resource budgets, which can be learned on the job and improve over time [13]. Rosha (2020) presented hybrid media models using a traditional modeling approach based on linear regressions based on data from a retailer's advertising scan, especially online and offline investments for each They examined online channels and conversion criteria. The results of the models indicate that online channels have been more effective in explaining the variance of the number of affiliates that represent sales. The results also show that certain factors for their personal activities affect the hybrid models of the media. It is expected that this accurate measurement in marketing and consequently the implementation of implemented media budgets and business performance [14].

Liu et al. (2020) examined budget efficiency optimization for online bidding advertising. They modeled this as a multi-constraint budget allocation optimization problem and used an innovative algorithm to approximate the budget allocation solution. They proposed a bidding strategy for filtering low-quality displays, which predicts PCTR rates. However, setting a pCTR threshold is challenging due to market uncertainty. They solved this problem by modeling the distribution of pCTRs and market prices over time. The results showed that the proposed method has the best performance in terms of different standard criteria (eg, number of clicks, cost per click) compared to other methods [15]. Suha (2020), in a study entitled "An optimization model to determine the appropriate budget for improving workplace safety" has presented an optimization model to determine the appropriate budget for improving workplace safety. The main factors of the model include a wide range of safety interactions and its implementation costs. The main purpose of this study is to create an optimization ideology to improve workplace safety. The optimization method developed in this study focuses on identifying the best set of safety countermeasures with limited budget. The first step in developing this approach was to budget appropriately, as engineers may need to know the appropriate budget to improve workplace safety at a construction project. The second or final task was to identify the best set of safety countermeasures in the budget. In this study, an optimization method in a work area was successfully developed and implemented as a case study. The study area was a 3-mile interstate highway construction project that began in May and ended in August. Previous studies have been used to identify a set of safety interactions that can be used for work area safety. This paper is also discussed to model the statistical methods used in the mathematical planning model [16]. Shi et al. (2020), in their research entitled "Joint optimization of budget allocation and maintenance planning of multi-facility transport infrastructure systems" to integrated budget allocation and the problem of preventive maintenance optimization for transport infrastructure systems that are deteriorating Is paid. They first developed an integer programming formulation for the problem and re-formulated the problem in order to solve large-scale problems, breaking it down into multiple models of the Markov decision-making process. They presented a two-step priority-based method for finding optimal maintenance decisions. Computational studies were performed to evaluate the performance of the proposed algorithms. The results show that the proposed algorithms are efficient and effective in finding suitable solutions for maintaining multi-facility systems. Also, by examining the characteristics of optimal maintenance decisions and making several important observations, it was shown that this approach can also be a useful decision-making guide for real-world problems [17].

Safari (2026) analyzed Airbnb reviews using NLP to examine how sentiment influences acceptance rates and pricing across U.S. regions. The study shows that sentiment polarity is more influential than review volume in shaping platform outcomes [18]. Jang (2019) proposed to provide a decision support framework for R&D budget allocation that maximizes overall R&D expectations. The proposed framework included a research and development output forecasting model with an optimization method. They first used a machine learning algorithm to accurately estimate future research and development efficiencies. Then, they used a robust optimization method to avoid uncertainty in the predicted R&D output values. The results provided insights for policymakers and researchers on better systems design and budgeting for national R&D programs [19]. Fingenshou et al. (2019) examined how the media influence agendas, resource allocation, and case decisions in public bureaucracies. They applied a hybrid approach to Norway using a comprehensive survey of government employees in ministries and agencies, as well as interviews with government employees and political leaders. The results clearly confirmed the notion that the media can influence agendas, resource allocation, and decision-making in ministries and organizations. Measures are taken when media pressure and widespread popular support increase, especially when issues are considered important to political actors, suggesting that when the media influences public bureaucracies. - Let, the forces under pressure and tension are also involved [20]. In the article Ponosami and Sodamati (2019), the focus is on the introduction, features, trends, applications and benefits of marketing using artificial intelligence. Based on a review of the literature on artificial intelligence and its applications in marketing, they offer the theory that artificial intelligence can be used to visualize future developments, in which users of digital economics use the applications of artificial intelligence in marketing as a tool for investment. Artificial intelligence has a dual role in the digital world and marketing economics and is used as a tool for the production of agencies, marketing agents and digital marketing. In this way, it can increase or decrease the level of profit through rational use of it [21].

In their 2019 paper, Rainers et al. primarily analyze the direct impact of the advertising brand industry on indirect advertising recall. Because the ultimate effectiveness of an ad depends on the interaction of a set of variables related to ad planning, the ad product, and the individual, rather than the individual impact of each of these variables alone; Therefore, this article examines the impact of the advertising brand industry on the relationship between the call and the following media planning variables: (1) ad duration, (2) brand advertising pressure, (3) the position of the ad in relation to the program, (4) Advertising clutter, (5) showing the duration of commercials, (6) the relative position of advertising in commercials, and (7) the effects of precedence and latency. Although several studies have examined the effects of each of the main planning variables separately, the main contribution of this paper is to analyze the relationship between the variables of advertising call and media planning from a comprehensive perspective. Achieving these goals provides a series of recommendations for optimizing television media planning depending on the advertising industry and provides the necessary empirical evidence for the advancement of knowledge and awareness in this field [22].

Wang et al. (2018) allocated a budget allocation in advertising media by presenting an integrated algorithm based on reinforcement training (RL) and multiple choice backpack problem (MCKP). In addition, they used some methods, such as cost discretization, to reduce model complexity. Also, the Q learning reward function was reconstructed by considering an additional impact factor between channels. Their experimental results showed more effectiveness than previous methods [23]. Paulson et al. (2018) presented a non-specific algorithm for the optimal allocation of Internet advertising budgets in the context of organized advertising. Unlike many DSP algorithms, which run each ad independently, this method explicitly calculates the correlation of visitors to websites. As a result, scan managers can make optimal decisions about all advertising opportunities. Most importantly, they can prevent publishers from over-pricing. The proposed method can also be used as a tool for budgeting. Because it easily provided optimal proposed guidelines for a wide range of scan budgets [24]. Ahmadi et al. proposed a computer vision method to detect crack on concrete [25]. They also used digital twin technology in the stormwater management and coastal conservation [26] [27]. Asgari et al. used data analysis of decision support for sustainable welfare in the presence of GDP threshold effects [28]. Moghim et al. research on integrated assessment of extreme hydrometeorological events in Bangladesh [29]. Safari and Kim (2026) proposed an Extreme Value Machine–based approach to enhance IoT security by detecting unknown and out-of-distribution attacks. Their results demonstrate improved resilience and robustness compared to traditional closed-set intrusion detection methods [30]. Danandeh Mehr et al. developed a new evolutionary explicit model for meteorological drought prediction at ungauged catchments [31] [32] [33]. Asadi et al. introduce a gradient based optimization techniques using multidimensional surface for 3D visualizations and initial point sensitivity [34].

## *2-1-Applications and Helping to Solve the COVID-19 Crisis*

One of the categories of Industry 4.0 is mobile technology, which is the title of the article due to its high importance. Mobile technologies, regardless of the type of operating system and features, are of great importance in people's daily lives. In the post-COVID-19 era, this technology is extremely practical and useful. Because the only safe way that can be connected to the whole world is to meet people's social needs, and the use of the facilities, sensors, and chips can also help control the crisis. Unquestionably, smartphones are based on these operating systems. Nowadays, different countries utilize new technologies in the medical industry. For example, Germany has implemented a tracking program based on smartwatches, which uses pulse and temperature and transmits the resulting data to areas of health bases for further analysis. This feature has been applied in a limited and experimental way to evaluate people with coronavirus disease. In this way, people are divided into organizational data according to the region. Due to some changes such as pulse and temperature in the reference program, the smartwatch owner may have coronavirus disease. In the same way, that area is under the control of the health staff. With this system, individuals' incidence and the speed of disease transmission can be significantly assessed. Every country takes the importance of mobile technologies seriously based on the number of casualties in this crisis by reviewing the release dates and launches of applications that contribute to the COVID-19 crisis. It can be seen that most of the countries that produce these applications started operating in "April and May ".In the meantime, only three countries, Austria, Singapore, and Israel, contributed to the COVID-19 crisis in their region with this technology in March and about a month and a half earlier than other countries.

**Table 1:** Some call tracking applications in different countries with the necessary permissions

| Call tracking app name | Country | Necessary permissions to run the program | Description |
|---|---|---|---|
| Ketju (Shen, Wei, & Li, 2020) | Finland | Bluetooth | The app works with Bluetooth to track people concerning their privacy to prevent coronavirus disease. |
| ViruSafe (Lalmuanawma et al., 2020; Virussafe, 2020) | Bulgaria | Internet, location, camera, flashlight, accounts | This program works through GPS and tries to prevent coronavirus disease in the population of the region. |
| CovTracer (Covtracer, 2020; Lalmuanawma et al., 2020) | Cyprus | Memory card, location, writing on the memory card, Internet, account synchronization | The app sends notifications to people who have been diagnosed with coronavirus with GPS tracking. |
| Immuni (Immuni, 2020; Lalmuanawma et al., 2020) | Italy | Bluetooth, location, internet access | This app does not collect any GPS-based data, such as private information.<br>The program works via Bluetooth and can also be used to communicate with a general practitioner in the case of acute symptoms. |
| HSE COVID-19 App (HSE.Covid19app, 2020) | Ireland | Bluetooth, view network connections | This app will get your medical records and information, control the people you have recently been in contact with by tracking your calls history, and notify you in case of suspicious symptoms. |
| CoronApp (CoronApp, 2020; Lalmuanawma et al., 2020) | Columbia | Access location, contacts, contact information, Bluetooth, prevent the device from falling asleep. | The program is completed with the public participation of individuals and alternately tries to provide solutions and send health aid by filling in information such as family members' symptoms. |
| StopCOVID (Lalmuanawma et al., 2020; StopCovid, 2020) | France | Access to location, camera, network, Bluetooth | This app warns you when you are near people with recent positive coronavirus test. While protecting the privacy of the user, the alarm is sent anonymously. |
| Mask.ir (Lalmuanawma et al., 2020; Mask.ir, 2020) | Iran | Vibration control, network access, run at startup, approximate and accurate locations based on network and GPS, camera, Bluetooth | In this program, you can see the map of coronavirus infection in your place of residence in terms of population and then complete the corona test information or find out through the program if you have been in contact with people who have had coronavirus. |
| Smittestopp (Lalmuanawma et al., 2020; Smittestopp, 2020) | Norway | Location access, Bluetooth | The program reminds you of the necessary health tips when you are near a person whose test is positive. |
| Ehteraz (Ehteraz, 2020; Lalmuanawma et al., 2020) | Qatar | Memory card access, delete or change memory card, direct call, location access, GPS, internet access, prevent the device from falling asleep. | Notifying and providing education and health tips to the people and prevention of COVID-19. |

As shown in Table 1, different countries have taken a step forward in software technologies to control diseases such as COVID-19. Each of them tries to control and evaluate the disease with a part of artificial intelligence divisions.

### *2-1-Technology in Production and Industry*

As the business relies more on technology, new production techniques such as 3D printers and industrial automation have increased in factories since the advent of the COVID-19. Moreover, import incentives have been declined. During this period, the use of smartphones became increasingly widespread and accelerated digital globalization. The global economy has been affected strongly by the outbreak of the corona, and industries such as tourism, aviation, industry, and manufacturing have been suffered in the three months of the disease since the supply chain management of industries such as automobiles, electronics, etc. faced problems with the closure of international flights. The lack of supply of spare parts worldwide is among them. The coronavirus had the most significant impact on the manufacturing industry in China, the United States, and Germany. As indicated in Table 2, if the virus is present in communities for a long time, the expert's knowledge becomes more about this virus because this virus is accompanied by alterations every moment and changes its symptoms. It can be said that the importance of artificial intelligence and digital style on the life of the industry is the only one industry the effects of which are visible in the short term, and there is no need for a long time to discover the truths.

**Table 2:** Applications of digitization if COVID-19 continues to be prevalent

| Period of time | Application | Description |
|---|---|---|
| Short-term (weekly) | Reduce virus outbreak | Without AI-related technologies, the chances of getting the virus are extremely high. |
| Medium-term (Monthly and Seasonal) | Diagnosis of the virus before the onset of serious symptoms | With city-wide image processing cameras, the city can be placed in demographic categories, and the symptoms of the people can be monitored, and the affected people can be immediately directed to the medical and health centers. |
| Long-term (Annual or more) | Diagnose, control, and improve patients or people at risk of catching the virus | With this disease's existence for a year, the symptoms of this disease have been placed in the form of a database and have been made available to all scientists, etc. All the ways of showing the symptoms are evaluated using this database. Also, it is introduced to the people in the form of devices and health-oriented tools. |

## 3. Methodology

### *3.1. The validity and reliability of research tools*

The purpose of the validity is that whether the measuring instrument can really measure the desired property. If the measuring device does not have sufficient validity in terms of the desired feature, the results would be worthless. In order to prevent it, the scientific validity of the questionnaire must first be achieved. The questionnaire was generally developed in two sections, Strategy and Management control system, based on theoretical studies. The Strategy section was originally designed and developed by developed by Jermias & Gani (2004). It is divided into two sections, Financial and Non-Financial factors, based on Firth (1996). Also, in the preparation and design of a questionnaire we used specialized experts in this field. Both questionnaires, divided into four categories, were tested by Cronbach's alpha validity test.

This research consists of five variables, which include the strategy that the company pursues. First the type of management control system that the company uses, which is also studied as an adjustment variable. Second the performance measurement of the company, which is the dependent variable of the research. The management control system is divided into two variables, the management control system based on financial factors and the management control system based on non-financial factors, by which 4 regression equations are as follows:

$$P = \beta 0 + \beta 1\ CLS + \beta 2\ FMCS + \beta 3\ CLS^{*} FMCS + \varepsilon \qquad (2)$$

$$P = \beta 0 + \beta 1\ DS + \beta 2\ NFMCS + \beta 3\ CLS^{*} NFMCS + \varepsilon \qquad (3)$$

$$P = \beta 0 + \beta 1\ CLS + \beta 2\ NFMCS + \beta 3\ CLS^{*}\ FMCS + \varepsilon \qquad (4)$$

$$P = \beta 0 + \beta 1\ DS + \beta 2\ FMCS + \beta 3\ DS^{*}\ FMCS + \varepsilon \qquad (5)$$

Used to test research hypotheses, all variables are present in regression equations, which include:

P: The company Performance is comparable to its leading competitor

CLS: The degree to which a company's strategy adapts to the cost leadership strategy

DS: The degree to which a company's strategy adapts to the differentiation strategy

FMCS: Compliance with the Financial Management Control System

NFMCS: Compliance with the Non-financial Control Management System

It is obtained by means of a questionnaire that has a value between values 1 to 5. The variables of this study, in addition to the performance measurement as dependent variable, other variables according to the hypothesis tested, will have the role of independent and control variables. Also, in the case of financial and non-financial Management control system variables, they will have the role of moderator variable.

### *3.2. Z-number Theory*

A Z-number is an ordered pair of $(\tilde{\boldsymbol{F}}, \tilde{\boldsymbol{L}})$, with $\tilde{\boldsymbol{F}}$ and $\tilde{\boldsymbol{L}}$ both being considered TFN. As a first component, $\tilde{\boldsymbol{F}}$ is a fuzzy subset of the Y domain. Z is a Z-number that is linked to a real-valued uncertain variable Y. $\tilde{\boldsymbol{L}}$ is a fuzzy subset of the generic interval [0, 1] as a second component. Where $\tilde{\boldsymbol{F}}$ refers for limitation and $\tilde{\boldsymbol{L}}$ stands for confidence or dependability, a Z-number may be employed to express information about an uncertain variable. The term "Z-information" refers to a collection of Z-values. In a lot of everyday thinking and decision-making, Z-information is used. Assume Y is a stochastic variable, based on the fuzzy constraint in Equation (2). The probabilistic constraint on Y is indicated by the probability distribution of Y. The following is the probabilistic limitation:

$R(Y){:}\, Y\ is\ p$ (6)

and the probability density function of Y is described in the (7) depending on Equation (6).

$$R(Y){:}\, Y\ is\ p \rightarrow prob\ (u \leq Y \leq u + du) = p(u)du \tag{7}$$

where $p$ is deferential of $u$ and $du$ denotes the probability density function of $Y$.

Suppose that $Z = [(a_1, b_1, c_1), (a_2, b_2, c_2)]$ to convert Z-number to TFN. The limitation is represented by the first component $(a_1, b_1, c_1)$, whereas dependability is represented by the second part $(a_2, b_2, c_2)$. The second element (reliability) is initially converted into a crisp integer, as follows:

$$\alpha = \frac{\int y\, \mu_{\tilde{L}}(y)dy}{\int\, \mu_{\tilde{L}}(y)dy} \tag{8}$$

Where $\mu_{\tilde{L}}(y)$ is the same as in Equation (2).

Then, $\alpha$ as the second element (reliability) contributes to the first part's weight (restriction).

Equation (9) may be used to derive the TFN form of the weighted Z-number:

$$\tilde{Z}' = (\sqrt{\alpha} * a_1, \sqrt{\alpha} * b_1, \sqrt{\alpha} * c_1) \tag{9}$$

## *2.3 Z-SWARA Method*

In a fuzzy context, fuzzy SWARA (fuzzy Step-wise Weight Assessment Ratio Analysis) is a multi-attribute decision-making approach for computing the weight of criterion and sub-criteria. The fuzzy SWARA method works in the same way as the SWARA method. However, due to uncertainty in decision-making or a lack of knowledge, it is expanded to fuzzy SWARA. In the fuzzy SWARA method, researchers play an important role in determining the weight of the criterion. Hence, the data collected depending on expert judgments. The fuzzy SWARA technique was expanded to the Z-SWARA technique in this work, and a reliability factor was introduced to increase the confidence of the outcome. The Z-SWARA technique is broken down into the following steps:

**Step 1**: Due to self, the professionals arrange criteria from most essential to least significant in decreasing order.

**Step 2**: Based on their first evaluation, experts must allocate linguistic characteristics to the relative significance of criterion $j$ about the previous $j - 1$ criteria. Experts then use Table 3 to calculate the value of the first component $(\tilde{F}_j)$. Table 2 is used to calculate the dependability component $(\tilde{L}_j)$. The result is a Z-number for each condition.

**Table 3:** For weighting criteria, linguistic characteristics are used

| Linguistic Variables | TFNs |
|---|---|
| Equally important (EI) | (1, 1, 1) |
| Moderately less important (MOL) | (2/3, 1, 3/2) |
| Less important (LI) | (2/5, 1/2, 2/3) |
| Very less important (VLI) | (2/7, 1/3, 2/5) |
| Much less important (MUL) | (2/9, 1/4, 2/7) |

**Table 4:** Variables in linguistics for assessing reliability

| Linguistic variables | Very Weak (VW) | Weak (W) | Medium (M) | High (H) | Very High (VH) |
|---|---|---|---|---|---|
| TFNs | (0,0,0.25) | (0.2,0.35,0.5) | (0.35,0.5,0.75) | (0.5,0.75,0.9) | (0.75,1,1) |

***Step 3***: The second component (reliability) is turned into a crisp number using Equation 1 to convert the Z-number produced in Step 2 to a TFN (6). The equation is then used to add the weight to the first component (7).

As an instance, assume the relative significance in the form of linguistic variables for the $j$-th criteria is $(VLI, M)$. The Z-number becomes $[(2/7,1/3,2/5),(0.35,0.5,0.75)]$ by changing the appropriate TFN values of $VLI$ and M from Tables 4 and 5, respectively. The Crisp value is $\alpha = 0.53$, and about Equation (7), the TFN form of Z-number is $(0.21,0.24,0.29)$. Table 5 shows more Z-number to TFN conversions.

**Table 5:** For weighting criteria, transformation procedures for Z-number to TFN depending on linguistic characteristics are used

| Linguistics Variables | TFNs | Linguistics Variables | TFNs |
|---|---|---|---|
| (EI,VW) | (1,1,1) | (LI,H) | (0.34,0.42,0.56) |
| (EI,W) | (1,1,1) | (LI,VH) | (0.38,0.48,0.64) |
| (EI,M) | (1,1,1) | (VLI,VW) | (0.08,0.10,0.12) |
| (EI,H) | (1,1,1) | (VLI,W) | (0.17,0.20,0.24) |
| (EI,VH) | (1,1,1) | (VLI,M) | (0.21,0.24,0.29) |
| (MOL,VW) | (0.19,0.29,0.43) | (VLI,H) | (0.24,,0.28,0.34) |
| (MOL,W) | (0.39,0.59,0.89) | (VLI,VH) | (0.27,0.32,0.38) |
| (MOL,M) | (0.49,0.73,1.10) | (MUL,VW) | (0.06,0.07,0.08) |
| (MOL,H) | (0.56,0.85,1.27) | (MUL,W) | (0.13,0.15,0.17) |
| (MOL,VH) | (0.64,0.96,1.44) | (MUL,M) | (0.16,0.18,0.21) |
| (LI,VW) | (0.12,0.14,0.19) | (MUL,H) | (0.19,0.21,0.24) |
| (LI,W) | (0.24,0.30,0.39) | (MUL,VH) | (0.21,0.24,0.27) |
| (LI,M) | (0.29,0.37,0.49) | | |

***Step 4:*** depending on the findings of Step 3, the fuzzy weight coefficient $\tilde{q}_j$ is defined as follows:

$$\tilde{q}_j = \frac{\tilde{q}_{j-1}}{\tilde{Z}'_j} \tag{10}$$

where $\tilde{q}_j$ is TFN and $\tilde{q}_1 = (1,1,1)$.

***Step 5:*** Ultimately, the relative weights of the $j-th$ assessment criteria were computed, taking into consideration $n$ assessment criteria:

$$\widetilde{w}_j = \frac{\tilde{q}_j}{\sum_{j=1}^{n} \tilde{q}_j} \tag{11}$$

where $\widetilde{w}_j$ is a TFN.

## *2.4. Z-WASPAS Method*

The fuzzy WASPAS approach [52], like the WASPAS methodology [50,51], is a multi-variable decision-making methodology for determining the system's certainty in highly sensitive situations. (See, for example, [53] on reservoir flood control administration.) (See, for example, [54] for the management of reservoir flood control.) The Weighted Sum Model (WSM) and the Weighted Product Model (WPM), two well-known MCDM techniques, are combined in the fuzzy WASPAS methodology (WPM). The fuzzy WASPAS approach must be used to analyze the favorable (e.g., profit, efficiency) or non-beneficial (e.g., cost) side of each risk element, which relies on expert judgments. Higher values are generally preferred for good features, while smaller values are typically desirable for non-beneficial features. Only the positive features of criteria are examined in this research, and the fuzzy WASPAS result is presented as a utility function ($K_i$) that may be used to rank alternatives. For rating failure modes, the new expanded Z-number of fuzzy WASPAS, called Z-WASPAS, is utilized. The Z-WASPAS technique is broken down into the following steps:

***Step 1***: For each piece, researchers firstly define a linguistic variable. Subsequently, to construct decision matrix H, the relevant value of each linguistic variable is provided to every element. Assume the Z-number $Z = (\tilde{F}_{ij}, \tilde{L}_{ij})$, which is the linguistic value for $\tilde{F}_{ij}$ in Table 4 [50]. $\tilde{L}_{ij}$can also extract linguistic values from Table 5 in an identical way as the Z-SAWARA technique. As a result, the following is the decision-making matrix H containing Z-number elements:

$$H = \begin{bmatrix} h_{11} & \dots & h_{1n} & .. \\ \dots & \dots & \dots & .. \\ h_{m1} & \dots & h_{mn} & .. \end{bmatrix} \quad (13)$$

The number of options is indicated by $h_{ij} = [(a_{ij}^f, b_{ij}^f, c_{ij}^f), (a_{ij}^l, b_{ij}^l, c_{ij}^l)], i = 1,..m, j = 1,..,n,$ $m$, and the number of criteria is shown by $n$.

**Table 4:** For rating failure modes, linguistic variables are used [50]

| **Linguistic variables** | **Very Poor (VP)** | **Poor (P)** | **Medium Poor (MP)** | **Fair (F)** | **Medium Good (MG)** | **Good (G)** | **Very Good (G)** |
|---|---|---|---|---|---|---|---|
| TFNs | (0,1,2) | (1,2,3) | (2,3.5,5) | (4,5,6) | (5,6.5,8) | (7,8,9) | (8,9,10) |

***Step 2:*** TFN ($\tilde{Z}'$) transforms the decision matrix H with Z-number elements. The following is the modified decision-making matrix H:

$$\tilde{\tilde{H}} = \begin{bmatrix} \tilde{\tilde{h}}_{11} & \dots & \tilde{\tilde{h}}_{1n} & .. \\ \dots & \dots & \dots & .. \\ \tilde{\tilde{h}}_{m1} & \dots & \tilde{\tilde{h}}_{mn} & .. \end{bmatrix} \quad (12)$$

where $\tilde{\tilde{h}}_{ij}$ is a TFN in the form of $\tilde{Z}'$, $i = 1,..m, j = 1,..,n, m$ represents the number of options, and n represents the number of criteria.

As an instance, imagine the expert determines the significance, $\tilde{F}_{ij}$, as "Medium poor" (MP) and the reliability, $\tilde{L}_{ij}$, as "Weak" (W) for the $i$-th choice and the $j$-th criterion. As a result, the Z-number and crisp value are $[(2,3.5,5), (0.20,0.35,0.50)]$, and $\alpha = 0.35$, respectively. As a result, the converted form of a Z-number ($\tilde{Z}'$) is $(1.18,2.08,2.96)$. Table 5 shows further Z-number to TFN ($\tilde{Z}'$) conversions.

wherein $\tilde{\tilde{h}}_{ij}$ is a TFN in the shape of $\tilde{Z}'$, $i = 1,..m$, $j = 1,..,n$, $m$ represents the number of options, and $n$ represents the number of criteria.

As an instance, imagine the professional determine the significance, $\tilde{F}_{ij}$, as "Medium poor" (MP) and the reliability, $\tilde{L}_{ij}$, as "Weak" (W) for the $i$-th choice and the $j$-th criterion. As a result, the Z-number and crisp value are $[(2,3.5,5), (0.20,0.35,0.50)]$, and $\alpha = 0.35$, respectively. As a result, the converted form of a Z-number TFN ($\tilde{Z}'$) is $(1.18,2.08,2.96)$. Table 5 shows further Z-number to TFN ($\tilde{Z}'$) conversions.

**Table 5:** For grading failure modes, transformation procedures to convert Z-number to TFN ($\tilde{Z}'$) depending on linguistic factors are used

| Linguistics Variables | TFNs | Linguistics Variables | TFNs |
|---|---|---|---|
| (VP,VW) | (0,0.29,0.58) | (F,H) | (3.39,4.23,5.08) |
| (VP,W) | (0,0.59,1.18) | (F,VH) | (3.83,4.79,5.74) |
| (VP,M) | (0,0.73,1.46) | (MG,VW) | (1.44,1.88,2.31) |
| (VP,H) | (0,0.85,1.69) | (MG,W) | (2.96,3.85,4.73) |
| (VP,VH) | (0,0.96,1.91) | (MG,M) | (3.65,4.75,5.84) |
| (P,VW) | (0.29,0.58,0.87) | (MG,H) | (4.23,5.50,6.77) |
| (P,W) | (0.59,1.18,1.77) | (MG,VH) | (4.79,6.22,7.66) |
| (P,M) | (0.73,1.46,2.19) | (G,VW) | (2.02,2.31,2.60) |
| (P,H) | (0.85,1.69,2.54) | (G,W) | (4.14,4.73,5.32) |
| (P,VH) | (0.96,1.91,2.87) | (G,M) | (5.11,5.84,6.57) |
| (MP,VW) | (0.58,1.01,1.44) | (G,H) | (5.93,6.77,7.62) |
| (MP,W) | (1.18,2.07,2.96) | (G,VH) | (6.70,7.66,8.62) |
| (MP,M) | (1.46,2.56,3.65) | (VG,VW) | (2.31,2.60,2.89) |
| (MP,H) | (1.69,2.96,4.23) | (VG,W) | (4.73,5.32,5.92) |
| (MP,VH) | (1.91,3.35,4.79) | (VG,M) | (5.84,6.57,7.30) |
| (F,VW) | (1.15,1.44,1.73) | (VG,H) | (6.77,7.62,8.47) |
| (F,W) | (2.37,2.96,3.55) | (VG,VH) | (7.66,8.62,9.57) |
| (F,M) | (2.92,3.65,4.38) | | |

***Step 3:*** Using the $\tilde{\tilde{H}}$ matrices to normalize the beneficial and non-beneficial components

$$\hat{h}_{ij} = \begin{cases} \dfrac{\tilde{\bar{h}}_{ij}}{\max\limits_i \tilde{\bar{h}}_{ij}} & for \quad j \qquad beneficial \\ and & \\ \dfrac{\min\limits_i \tilde{\bar{h}}_{ij}}{\tilde{\bar{h}}_{ij}} & for \quad i \qquad non-beneficial \end{cases} \tag{14}$$

***Step 4:*** Compute the weighted normalized fuzzy decision-making matrix of $\hat{h}_{ij}$ for the WSM ($\tilde{Q}$ TFN) and WPM ($\tilde{P}$ TFN) as follows:

$$\tilde{Q}_i = \sum_{j=1}^{n} \hat{h}_{ij}\tilde{w}_j \tag{15}$$

$$\tilde{P}_i = \prod_{j=1}^{n} \hat{h}_{ij}^{\tilde{w}_j} \tag{16}$$

Evaluate the center-of-area of each TFN for decision making to assist in de-fuzzify performance assessment [49]:

$$\bar{Q}_i = \frac{1}{3}(a_i^Q + b_i^{Q_i} + c_i^Q) \tag{17}$$

$$\bar{P}_i = \frac{1}{3}(a_i^P + b_i^P + P_i^c) \tag{18}$$

***Step 5:*** For the $i$-th choice, use the utility function $K_i$ to determine the value of failure modes and rank them as follows:

$$K_i = \lambda \sum_{j=1}^{m} \bar{Q}_i + (1-\lambda) \sum_{j=1}^{m} \bar{P}_i \,; \quad 0 \le \lambda \le 1, \tag{19}$$

$$0 \le K_i \le 1,$$

where, $\lambda = \frac{\sum_{i=1}^{m} \bar{P}_i}{\sum_{i=1}^{m} \bar{Q}_i + \sum_{i=1}^{m} \bar{P}_i}$.

Ultimately, the options may be sorted from highest to lowest $K_i$ value.

# 4. Results and Discussion

Cronbach's alpha coefficient was employed in this research to pre-test the questionnaire's reliability in terms of the link between business strategy, management control system, and business unit performance assessment. Prior to the final implementation, 29 research samples were randomly selected and a questionnaire was provided to them. Using the data obtained from these questionnaires and using SPSS, the confidence factor was calculated using Cronbach's alpha method. The questionnaire data is generally divided into four categories of questions; All four categories, as seen in their reliability tables, have significant internal correlations between questions. This is because the Cronbach's alpha is significantly higher than 0.7 among all four categories of questions. Therefore, questions can be assessed on the validity and reliability. The total amount of Cronbach's alpha for each of the four categories of relevant items and the overall correlation of each item, also, the amount of Cronbach's alpha after deleting each item are listed in Tables 6 to 9.

**Table 6.** Reliability test related to strategy

| | **Cronbach's Alpha if Item Deleted** | **The overall correlation of each item with the overall scale** |
|---|---|---|
| Product price | 0.897 | 0.587 |
| Research and Development for sale | 0.875 | 0.862 |
| Company brand | 0.879 | 0.822 |
| Activities related to product development | 0.888 | 0.709 |
| Program change rate | 0.910 | 0.378 |
| Product Distribution Standards | 0.890 | 0.693 |
| Product quality | 0.894 | 0.635 |
| After-sales service | 0.895 | 0.635 |
| Product Features | 0.882 | 0.776 |
| General reliability | 0.901 | 0.584 |

In order to test the effect of the Management control system moderator on the relationship between performance and strategy, moderated regression analysis has been used. In addition to the independent variable (strategy) and the moderator variable (Management control system), there is a multiplicity of these two variables. In the regression output, $R^2$ represents what ratio of total variability or variance in the dependent variable is explained by the independent variables.

**Table 7:** Reliability test related to performance measurement

| | **Cronbach's Alpha if Item Deleted** | **The overall correlation of each item with the overall scale** |
|---|---|---|
| Capital return rate | 0.924 | 0.701 |
| Profitability | 0.915 | 0.851 |
| Operating cash flow | 0.921 | 0.784 |
| Cost control | 0.37 | 0.454 |
| Development of new products | 0.926 | 0.644 |
| Sales volume | 0.918 | 0.796 |
| Market share | 0.915 | 0.852 |
| Market development | 0.917 | 0.809 |
| human recourse development | 0.917 | 0.819 |
| General reliability | 0.930 | |

**Table 8:** Reliability test related to non-financial factors of management control system

| | Cronbach's Alpha if Item Deleted | The overall correlation of each item with the overall scale |
|---|---|---|
| Customer satisfaction | 0.904 | 0.507 |
| Timely delivery | 0.891 | 0.643 |
| Reliable delivery | 0.891 | 0.643 |
| Criteria for determining key products | 0.880 | 0.751 |
| Quality | 0.886 | 0.690 |
| Test | 0.873 | 0.815 |
| Employee-based actions | 0.888 | 0.672 |
| strategic planning | 0.876 | 0.807 |
| General reliability | 0.899 | |

**Table 9:** Reliability test related to financial factors of management control system

| Cronbach's Alpha | items |
|---|---|
| 0.855 | 0.23 |

If $R^2$ is equal to 1, it indicates that the dependent variable is completely predictable through independent variables. A value of zero indicates that the dependent variable is not linearly related to the independent variables. If the regression coefficients of the multiplier variable are significant, the modification effect of the moderator variable can be inferred.

In the first part of the whole data, the entered sample is divided into two groups:

- Consists of people who have been looking for a differentiation strategy.
- Consists of people who are looking for a cost leadership strategy.

According to the paper (Tsamenya et al. 2011), people with a strategy value less than (3 mean values) are considered as people under the leadership strategy, and people who have a strategy value of 3 or more are considered as people under differentiation strategy. The moderated regression for each category was performed separately. For both regression categories, performance measurement is considered as a dependent variable. In the leadership group, the cost of strategic value of individuals is less than 3 and the absolute value of the score is considered as the value of strategy of individuals, because in the strategy variable, lower values mean more reliance on cost leadership strategy. An approximate value of 3 will indicate the dependence of at least on the cost leadership strategy. The independent variable strategy and Management control system based on financial factors will be the cost variable for the leadership, and for the group pursuing the differentiation strategy, the Management control system adjustment variable will be based on non-financial factors. The hypotheses after the correlation test to examine the relationship between the studied variables have been investigated by controlling other variables involved in the relationship by moderated regression analysis. In this analysis, apart from the independent variable and the modifier variable, the interaction product of the independent variable and the modifier variable in the regression equation will be considered.

Because the distribution of this data is not normal and the variance of the different groups is not equal, we can test the non-parametric Kruskal–Wallis one-way analysis of variance between each of the elements of strategic position, performance measurement and financial and non-financial management control system. The results of Moderated regression ($R^2$) analysis between variables show that there is a significant amount of significant relationship between business strategy variables and performance measurement. There is also the least amount of semantic relationship between the variables of financial management system and non-financial management system (Table 8). Considering that in this test, performance measurement is as a dependent variable and the other three variables are as independent variables. Therefore, we can conclude that among the independent variables, the most correlation is related to the business strategy variable and the lowest is related to the management control system-non-financial variable. Considering that the adjusted coefficient in most of these relations is close to one, it can be stated that there is a non-linear relationship between the independent variables and the dependent variable and the effect of modifying variable can be deduced.

**Table 10:** Moderated regression($R^2$) analysis between variables participating in the test

| variables | business strategy | management control system- financial | management control system- non-financial | performance measurement |
|---|---|---|---|---|
| business strategy | 1 | | | |
| management control system- financial | 0.852** | 1 | | |
| management control system- non-financial | 0.721* | 0.657* | 1 | |
| performance measurement | 0.928** | 0.869** | 0.707* | 1 |

* Significance at 95% confidence level

** Significance at 99% significance level

The results obtained from the Kruskal-Wallis test show that all variables have a significant difference of 99%; then the hypothesis is rejected (Table 9). Accordingly, to observe the different levels of factors within each component, we will go to the average ranking

**Table 11:** Analysis of Kruskal-Wallis test in business strategy variables, management control system (financial-non-financial) and performance measurement

| variables | business strategy | management control system- financial | management control system- non-financial | performance measurement |
|---|---|---|---|---|
| Chi-square | 8.702 | 9.534 | 9.351 | 7.421 |
| Df | 8 | 8 | 7 | 2 |
| | 0.005 | 0.001 | 0.006 | 0.000 |

The average ranking in the business strategy shows that the Product quality factor is at the highest level with 4.02 and the After-sales service factor is at the lowest level with 2.68. In the average business strategy, it has been observed that the factors of Product price, Company brand, Activities related to product development, Program change rate, After-sales service and Product Features with an average of less than 3 are considered as leadership strategies. On the other hand, factors such as Research and Development for sale, Product Distribution Standards and Product quality are called as differentiation strategies. (Table 10).

**Table 12:** Mean of ranking in tested dimensions in business strategy variable based on Kruskal-Wallis test

| Business strategy variable | Mean of ranking |
|---|---|
| Product price | 2.87 |
| Research and Development for sale | 3.21 |
| Company brand | 2.95 |
| Activities related to product development | 2.74 |
| Program change rate | 2.84 |
| Product Distribution Standards | 3.01 |
| Product quality | 4.02 |
| After-sales service | 2.68 |
| Product Features | 2.91 |

In the results of the average ranking, the Development of new products performance component is at the highest level and the sales volume is at the lowest level (Table 11). In the non-financial management control system component, the Customer satisfaction factor and in the financial component, Cost standards are at the highest level. Also, the strategic planning factor with a rate of 2.95 and the Deviation analysis factor with a rate of 2.87 are at the lowest level, respectively. (Tables 12 and 13).

**Table 13:** Mean of ranking in tested dimensions in performance measurement variable based on Kruskal-Wallis test

| Performance measurement variable | Mean of ranking |
|---|---|
| Capital return rate | 3.95 |
| Profitability | 2.54 |
| Operating cash flow | 2.98 |
| Cost control | 3.05 |
| Development of new products | 3.84 |
| Sales volume | 2.46 |
| Market share | 2.59 |
| Market development | 2.91 |
| human recourse development | 3.43 |

**Table 14:** Mean of ranking in tested dimensions in management control system- to non-financial variable based on Kruskal-Wallis test

| Management control system- to non-financial | Mean of ranking |
|---|---|
| Customer satisfaction | 4.26 |
| Timely delivery | 4.08 |
| Reliable delivery | 3.42 |
| Criteria for determining key products | 3.26 |
| Quality | 3.02 |
| Test | 3.37 |
| Employee-based actions | 3.29 |

| strategic planning | 2.95 |
|---|---|

**Table 15:** Mean of ranking in tested dimensions in management control system- to financial variable based on Kruskal-Wallis test

| Management control system- to financial | Mean of ranking |
|---|---|
| Cost standards | 3.30 |
| Deviation analysis | 2.87 |
| Budget control | 3.06 |

Our results show that factors such as product distribution standards and product quality as part of the differentiation strategy are similar to factors in the financial management control system such as customer satisfaction, quality and criteria for determining key products and they are consistent with the findings of Langfield and Smith, 1997 states that the non-financial management control system is consistent with the differentiation strategy.

## 5. Results of Failure modes prioritization

When utilizing the traditional FMEA approach, one of the primary issues that decision-makers encounter is assigning identical ratings to distinct failure scenarios. Because of a lack of organizational resources, decision-makers are unable to effectively identify important failure modes and take remedial action to mitigate their negative impacts. By creating this traditional technique relying on the SWARA and WASPAS methodologies and employing the Z-number concept, this research attempted to provide the findings with great separability compared to the FMEA. Furthermore, as compared to the fuzzy WASPAS technique, the Z-WASPAS technique may produce more realistic findings by combining the notions of uncertainty and dependability. In sum, F2 is the primary cause of risk in businesses, according to FMEA methodology and the opinions of three professionals. Modifications in need as a result of a weaker economy or other macro-economic modifications, adjustments in the cost of raw materials critical to Balco's manufacturing, and alteration in competition or pricing by rivals are all examples of industry and market risks. F6 or financial challenges with cash flow, capital, or cost constraints is recognized as a major failure mode in rank 2 in the suggested technique, which involves reliability aspects in detecting failure modes. When the quantity of money pouring into a firm doesn't satisfy the standards for accounts payable, it's a problem. Cash flow is the lifeblood of every successful business, and if it isn't adequate, you might be on the verge of going bankrupt.

Finding a flexible line of credit that allows businesses to access money when they're needed might be a simple method to weather a cash flow storm. The findings also show that F3, or fluctuations among consumers or in demand, is the most common failure mechanism. The recommended remedial action is to Demographics and consumer tastes are rapidly changing, making it more difficult than ever to keep up with client expectations. Globally, rising incomes, middle-class bulges, aging populations, and next-generation Millennials are reshaping the customer base. Consumer expectations for tailored experiences have risen as a result of technological advancements. Customers have grown used to customized, rapid service at cheap costs because to internet streaming, Uber, Amazon Prime, and other services.

**Table 16:** compared to existing strategies, the suggested methodology prioritizes failure modes

| | |
|---|---|
| mergers, acquisitions, and other competition. | **$F_1$** |
| market or industry changes. | **$F_2$** |
| changes among customers or in demand. | **$F_3$** |
| change management. | **$F_4$** |
| human resource issues, such as staffing. | **$F_5$** |
| financial issues with cashflow, capital or cost pressures. | **$F_6$** |
| IT disasters and equipment failure. | **$F_7$** |
| Planned system changes | **$F_8$** |
| Intermediaries and interfaces of facilities and devices | **$F_9$** |

**Table 17**: compared to existing strategies, the suggested methodology prioritizes failure modes

| **Failure Modes** | **Conventional FMEA** | | **Fuzzy-WASPAS** | | **Z-WASPAS** | |
|---|---|---|---|---|---|---|
| | **RPN** | **Rank** | $K_i$ | **Rank** | $K_i$ | **Rank** |
| **$F_1$** | 4660 | 6 | 0.817 | 6 | 0.74 | 6 |
| **$F_2$** | 8694 | 2 | 1.094 | 2 | 0.90 | 2 |
| **$F_3$** | 7903 | 3 | 1.065 | 3 | 0.89 | 3 |
| **$F_4$** | 4893 | 5 | 0.758 | 5 | 0.85 | 4 |
| **$F_5$** | 4578 | 7 | 0.920 | 4 | 0.51 | 9 |
| **$F_6$** | 8752 | 1 | 1.094 | 1 | 1.00 | 1 |
| **$F_7$** | 3235 | 8 | 0.470 | 9 | 0.68 | 8 |
| **$F_8$** | 2100 | 9 | 0.585 | 8 | 0.72 | 7 |
| **$F_9$** | 5679 | 4 | 0.697 | 7 | 0.76 | 5 |

## *5.1. Results of Sensitivity Analysis*

In five distinct scenarios, a sensitivity analysis is performed by altering the weight values of criterion (see Table18). The basic crisp weights of the criteria computed using the Z-SWARA technique in this study are represented by Case 0. The crisp value of weights are provided to the SODCT components and Case 1 to Case 4 are produced to evaluate how the rank of alternatives changes in potential situations. Table 19 shows the results of a sensitivity analysis for rating the outcomes of ten failure types and various instances. The order of significance in SODCT factors is S, C, D, T, and O, according to the aggregated judgment of three sets of decision-makers in this research. In compared to component C, factor S has a substantial influence on ME regulation. In all situations, F 2 is the failure mode with the greatest risk priority, based on Table 19. In terms of SODCT factors, F6 is the second most significant failure mode in Case 2, Case 3, and Case 4 due to the high weight of S, but in Case1and Case 4, F 3 is the second most essential failure mode due to the lower weight of S. This comparison holds true for a variety of additional criteria and failure types as well. According to the results of the sensitivity analysis, the weight of criterion has a substantial impact on the final ranking orders of failure modes.

**Table 18:** Weights and crisp value of weights of SODCT factors in different cases

| Risk factor | Case 0 | Case 1 | Case 2 | Case 3 | Case 4 |
|---|---|---|---|---|---|
| $w_S$ | (0.26,0.30,0.35) | (0.38,0.32,0.48) | (0.22,0.23,0.39) | (0.46,0.28,0.38) | (0.02,0.01,0.09) |
| $\alpha_{w_S}$ | 0.234 | 0.5 | 0.28 | 0.46 | 0.08 |
| $w_O$ | (0.36,0.32,0.38) | (0.22,0.29,0.32) | (0.16,0.10,0.19) | (0.15,0.01,0.17) | (0.20.12,0.36) |
| $\alpha_{w_O}$ | 0.363 | 0.2 | 0.05 | 0.02 | 0.19 |
| $w_D$ | (0.25,0.12,0.29) | (0.26,0.15,0.31) | (0.29,0.23,0.32) | (0.12,0.08,0.14) | (0.20,0.19,0.34) |
| $\alpha_{w_D}$ | 0.115 | 0.2 | 0.28 | 0.11 | 0.29 |
| $w_C$ | (0.25,0.32,0.36) | (0.02,0.01,0.13) | (0.09,0.06,0.16) | (0.39,0.20,0.29) | (0.35,0.25,0.38) |
| $\alpha_{w_C}$ | 0.185 | 0.05 | 0.11 | 0.29 | 0.29 |
| $w_T$ | (0.09,0.15,0.22) | (0.09,0.03,0.19) | (0.28,0.33,0.39) | (0.10,0.09,0.28) | (0.19,0.20,0.29) |
| $\alpha_{w_T}$ | 0.103 | 0.05 | 0.28 | 0.12 | 0.15 |

**Table 19:** In order to rank the failure modes' results in relation to the various instances

| Failure Modes | Case $_0$ | Case $_1$ | Case $_2$ | Case $_3$ | Case $_4$ |
|---|---|---|---|---|---|
| **$F_1$** | 8 | 6 | 9 | 8 | 7 |
| **$F_2$** | 1 | 1 | 1 | 1 | 1 |
| **$F_3$** | 3 | 2 | 4 | 5 | 2 |
| **$F_4$** | 4 | 5 | 8 | 3 | 6 |
| **$F_5$** | 8 | 9 | 3 | 7 | 4 |
| **$F_6$** | 5 | 7 | 2 | 2 | 2 |
| **$F_7$** | 9 | 8 | 6 | 1 | 5 |
| **$F_8$** | 2 | 4 | 6 | 9 | 1 |
| **$F_9$** | 6 | 4 | 7 | 4 | 9 |
| **$F_{10}$** | 7 | 9 | 5 | 6 | 8 |

As a result, choosing the appropriate weight for criteria based on the actual scenario is critical to risk prioritizing of failure modes and subsequent remedial measures. As previously mentioned, this research tried to offer an expanded methodology for prioritizing ME failure scenarios utilizing FMEA, SWARA, and WASPAS methodologies. In compared to the FMEA method, this method included additional cost and time variables depending on the case study in addition to standard ones. Furthermore, utilizing the newly established Z-SWARA technique, this study attempted to give alternative weights to risk variables. When compared to other traditional techniques, such as the AHP method, this technique uses less pairwise comparisons, and this study took into account the notions of uncertainty and dependability in the process of calculating the weights of risk variables concurrently utilizing the Z-SWARA technique. Other techniques, such as AHP or ANP, build models depending on criteria and expert assessments that influence priorities and rankings. As a result, SWARA can be beneficial for specific challenges where the priorities are known ahead of time based on the circumstance [57].

Furthermore, compared to the standard RPN score, the WASPAS technique centered on the Z-number concept has been developed and used in the suggested methodology, resulting in a more distinct prioritizing of failures. Growing market innovations suggest that the life of new items is shortening and design expenses are rising. As a result, before embarking on an expensive design and creating a production team, each firm must investigate the viability of product manufacturing and its potential to create the required profit. Globalization, outsourcing, and the establishment of strategic alliances are among the numerous alterations and revolutions that are taking place in the world of business and industry.

## 6. Conclusion

The proliferation of complex online media has accelerated the process of ideology. The formation process can also be influenced by interested people through advertising channels, the media through which messages are distributed to the target audience. Political construction campaigns and marketing departments split their advertising budgets between these channels (e.g., TV ads, website banner ads, billboards) to the ultimate goal (e.g., votes, sales) to the maximum. Therefore, the study of the problem of media resource allocation is significant in this discussion. In particular, the mechanisms of influence of opinion can be classified into two types based on its direct origin. First, there are endogenous influence mechanisms (e.g., word of mouth), in which individuals process the opinions expressed by other people they meet and validate their level of familiarity and trust in the new opinion. Based on information considered. This leads to the concept of endogenous effect weight diagram that attracts endogenous effect between individuals. On the other hand, there are mechanisms of exogenous influence, in which an external influence seeks to shape an individual's views. This mechanism is facilitated by various advertising channels. Finding the optimal budget allocation is complicated by several factors:

(1) Access to each channel is limited and there are significant overlaps between the target audiences of different channels.

(2) Different channels have different costs and trying to influence the ideas of external sources - can affect people in different ways and sometimes vice versa.

(3) The budget allocation decision is dynamic (time dependent) and changes with network status.

In addition, this influencer faces several deals: Early use of an advertising channel allows affected people to pass on the influence to their neighbors. There is also a trade-off between using cheap channels versus using expensive but effective channels. These rival forces make it difficult to determine the optimal allocation of funds. There are also important technical challenges to solving this problem, because describing the optimal budget allocation between channels over a period of time requires describing the optimal finite vector structure of the controls on the chart.

To achieve the most productivity in marketing, you need to determine how a given budget can be allocated to all channels to do business-oriented actions such as revenue, ROI (return on investment), growth rate. And to maximize like it. To do this, we use media mix modeling. The motivation for this study is, first, to prove that the proposed algorithm for allocating media budgets works in marketing and advertising campaigns, and second, to set up a framework in which real information can be processed, and the result They are available quickly.

In this paper, we presented a systematic review to evaluate the association between business strategy and management control system and their impacts on financial performance measurement of a production organization. In order to achieve the conceptual model, we used the questionnaire method with 185 employees randomly. Then 116 satisfactory questionnaires were chosen due to the incompleteness of some questionnaires. In this condition, the sample error level is incremented by 0.018. Accordingly, the size of the chosen final individual is 116 people, whose error level for the unknown community is 0.088. This research includes five variables, which include the policy that a company continues. First, the type of management control system that the business uses, which is also analyzed as an improvement variable. Second the performance measurement of the business, which is the dependent variable of the study. The management control system is divided into two variables, the management control system based on financial factors and the management control system based on non-financial factors. The variables of this paper, in addition to the performance measurement as a dependent variable, other variables according to the hypothesis tested, will have the role of independent and control variables. Also, in the case of financial and non-financial Management control system variables, they will have the role of the moderator variable. In the results of the average ranking, the Development of a new product performance element is at the most important level and the sales quantity is at the most under level. In the non-financial management control system component, the Customer satisfaction factor, and in the financial component, Cost standards are at the essential level. Also, the strategic planning factor with a rate of 2.95 and the Deviation analysis factor with a rate of 2.87 is at the lowest value, respectively. Our results show that factors such as product distribution standards and product quality as part of the differentiation strategy are similar to factors in the financial management control system such as customer satisfaction, quality, and criteria for managing key products and they are consistent with the findings of Langfield and Smith, 1997 states that the non-financial management control system is compatible with the differentiation approach. Financial risk analysis and control has grown more essential in the activities of commercial and non-commercial organizations. As a result, in a competitive market, businesses must place a high priority on strategic cost control in order to stay afloat. Also, In this case, a decision-making strategy centered on the FMEA is used to identify and prioritize risk factors financial of the control system in companies. Nevertheless, because this strategy has some significant limitations, this research has presented a decision-making approach depending on Z-number theory. For tackle some of the RPN score's drawbacks, the suggested decision-making methodology combines the Z-SWARA and Z-WASPAS techniques with the FMEA method.

The findings of this study, which involved applying and comparing the suggested method with the traditional FMEA and fuzzy-WASPAS techniques, showed that prioritizing failure modes using the suggested approach is closer to reality due to the incorporation of dependability factors. In sum, market or industry changes is the primary cause of risk in businesses, according to FMEA methodology and the opinions of three professionals. Decision-makers, on the other hand, can offer a variety of appropriate corrective/preventive measures for substantial failures, carry out the corrective actions with the appropriate departments, and evaluate the new circumstances and the effectiveness of the measures. The study's main flaw is that it fails to notice the cause-and-effect link between failure modes.